\documentclass[
 preprint, amsmath,amssymb,
 aps, physrev]{revtex4-2} % linenumbers

\usepackage{graphicx}% Include Figure files
\usepackage{dcolumn}% Align table columns on decimal point
\usepackage{bm}% bold math
\usepackage{xcolor}
\usepackage{hyperref}% add hypertext capabilities
\usepackage{subfig}

\begin{document}

\preprint{}

\title{\textbf{From Qubits to Couplings: A Hybrid Quantum Machine Learning Framework for LHC Physics} 
}

\author{Marwan Ait Haddou}
\email{marwan.aithaddou@edu.uca.ac.ma}
 \affiliation{LPMC, Faculty of Sciences Ben M’sick, Hassan II University of Casablanca, Morocco.}

\author{Mohamed Belfkir}%
 \email{Contact author: m\_belfkir@uaeu.ac.ae}
\affiliation{%
Department of Physics, United Arab Emirates University, Al-Ain, UAE
}%

\author{Salah Eddine El Harrauss}
\email{selharrauss@uaeu.ac.ae}
\affiliation{
Department of Physics, United Arab Emirates University, Al-Ain, UAE
}

\date{\today}% It is always \today, today,
\begin{abstract}
In this paper, we propose a new Hybrid Quantum Machine Learning (HyQML) framework to improve the sensitivity of double Higgs boson searches in the \( HH \to b\bar{b}\gamma\gamma \) final state at $\sqrt{s} = 13.6~\text{TeV}$. The proposed model combines parameterized quantum circuits with a classical neural network meta-model, enabling event-level features to be embedded in a quantum feature space while maintaining the optimization stability of classical learning. The hybrid model outperforms both a state-of-the-art XGBoost model and a purely quantum implementation by a factor of two, achieving an expected 95\%~CL upper limit on the non-resonant double Higgs boson production cross-section of \(1.9\times\sigma_{\text{SM}}\) and \(2.1\times\sigma_{\text{SM}}\) under background normalization uncertainties of 10\% and 50\%, respectively. In addition, expected constraints on the Higgs boson self-coupling $\kappa_{\lambda}$ and quartic vector-boson–Higgs coupling $\kappa_{2V}$ are found to be improved compared to the classical and purely quantum models.

\end{abstract}

\maketitle

%\tableofcontents

\section{Introduction}
\label{sec:intro} 

The discovery of the Higgs boson in 2012 by the ATLAS and CMS experiments at the Large Hadron Collider (LHC)~\cite{Aad_2012,Chatrchyan_2012} marked a cornerstone in the validation of the Standard Model (SM) of particle physics. Over the past decade, an extensive program of precision measurements has established the properties of the Higgs boson including its mass, spin and couplings to fermions and gauge bosons to be consistent with SM predictions within current experimental uncertainties~\cite{Higgs_mass_ATLAS,Higgs_mass_CMS}.  
However, the shape of the Higgs potential and in particular the strength of the trilinear Higgs self-coupling $\lambda_{3}$ remains largely unconstrained. This coupling governs the dynamics of electroweak symmetry breaking and the stability of the Higgs vacuum. Deviations from its SM expectation value, commonly expressed via the modifier $\kappa_{\lambda}=\lambda_{3}^{\text{BSM}}/\lambda_{3}^{\text{SM}}$, would provide direct evidence of an  extended scalar sector or new physics beyond the SM (BSM)~\cite{PhysRevD.110.030001,ATLAS:2022jtk,CMS:2024awa}.

The most sensitive process for probing $\kappa_{\lambda}$ at the LHC is Higgs boson pair production ($pp\to HH$), which proceeds mainly through gluon--gluon fusion (ggF) with a subdominant contribution from vector-boson fusion (VBF).  
Among the possible decay channels of the Higgs pair, $HH\to b\bar{b}\gamma\gamma$ offers an optimal balance between a clean experimental signature, excellent di-photon mass resolution and manageable backgrounds, despite its small branching fraction of approximately 0.26\% for $m_H=125~\text{GeV}$~\cite{CERNYellowReportPageBR}.  
Recent ATLAS and CMS searches have set expected upper limits on the non-resonant double Higgs cross-section at approximately three to five times the SM prediction, leaving significant room for methodological improvements to enhance sensitivity~\cite{atlas_results,CMS:2024awa}.

Machine learning (ML) techniques have become indispensable in collider physics enabling efficient discrimination between signal and background by exploiting complex high-dimensional correlations among observables~\cite{hepmllivingreview,Duarte:2024lsg}. Classical ML approaches such as tree-based algorithms, convolutional neural networks or graph neural networks have delivered substantial gains in searches for new physics~\cite{Belfkir:2025zlx}.  
Nevertheless, these methods remain bounded by classical computational architectures and feature representations, motivating the exploration of novel paradigms that may provide a representational advantage.

In this context, \emph{Quantum Machine Learning} (QML) has emerged as a promising new direction leveraging quantum computation to process information in exponentially large Hilbert spaces~\cite{Guan:2020bdl}. QML operates on quantum bits (qubits) instead of classical bits allowing the encoding of data into quantum states and the execution of transformations that can naturally capture high-order correlations in complex datasets. For example, the Quantum Support Vector Machine (QSVM)~\cite{Havlicek:2018nqz,Schuld:2018ahn} which employs quantum kernel estimation to achieve accurate classifications for certain benchmark datasets.  
With rapid advances in superconducting and photonic quantum hardware the feasibility of deploying QML algorithms on noisy intermediate-scale quantum (NISQ) devices has become increasingly realistic \cite{Arute:2019zxq,Guan:2020bdl}.  

Proof-of-principle studies in high-energy physics have already demonstrated the applicability of QML to detector simulation, particle reconstruction, and signal--background separation tasks~\cite{Wu:2021xsj,Fadol:2022umw}.  
Recent implementations of the QSVM-Kernel algorithm on both quantum simulators and superconducting hardware have achieved performance comparable to classical classifiers on small-scale datasets despite being limited by hardware noise, circuit depth and scalability~\cite{Wu:2021xsj}. These constraints highlight the current gap between the theoretical potential of fully quantum algorithms and their practical deployment on near-term devices. To bridge this gap, this study introduces a \emph{Hybrid Quantum Machine Learning} (HyQML) model combining parameterized quantum circuits with classical neural networks that exploits the representational power of quantum state spaces while retaining the optimization stability and scalability of classical learning frameworks. Such hybrid architecture provides a promising pathway toward realizing quantum advantage in realistic particle physics analyses.

In this work, we demonstrate the potential of such HyQML to improve the sensitivity to the Higgs boson self-coupling in the $HH\to b\bar{b}\gamma\gamma$ channel at $\sqrt{s}=13.6~\text{TeV}$.  
Using the same simulated datasets and event selections as in our previous study~\cite{Belfkir:2025zlx}, we develop a hybrid quantum--classical architecture that maps event-level features into a quantum feature space through parameterized quantum circuits.  
The objective is to assess whether such hybrid models can yield improved signal significance and tighter constraints on $\kappa_{\lambda}$ compared to established classical classifiers such as the XGBoost described in our previous work~\cite{Belfkir:2025zlx}, while simultaneously probing the prospects of quantum-enhanced learning for collider physics applications.

The rest of this paper is organized as follows: 
Section~\ref{sec:HH_intro} provides an overview of Higgs boson pair production at the LHC.  
The Monte Carlo (MC) simulation setup and generated samples are described in Section~\ref{sec:MC}.  
Section~\ref{sec:selection} details the physics object definitions and event selection criteria used in the analysis.  The HyQML framework including model architecture and training procedure is presented in Section~\ref{sec:hyQML}. The resulting classification performance, statistical interpretation and constraints on the Higgs self-coupling are discussed in Section~\ref{sec:results}, and the main conclusions are summarized in Section~\ref{sec:conclusion}.

\section{Higgs boson pair production}
\label{sec:HH_intro}

\begin{figure}[ht]
    \centering
    \subfloat[a][ggF box]   {\includegraphics[width=0.33\linewidth]{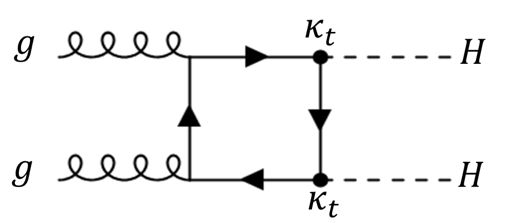}}\hspace{0.2cm}
    \subfloat[b][ggF triangle]{\includegraphics[width=0.35\linewidth]{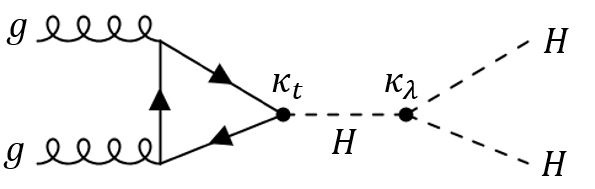}}\\
    \subfloat[c][VBF $\kappa_{\lambda}$]{\includegraphics[width=0.33\linewidth]{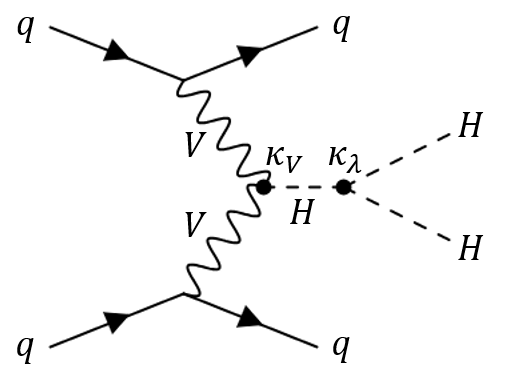}}\hspace{0.2cm}
    \subfloat[d][VBF $\kappa_{V}$]{\includegraphics[width=0.28\linewidth]{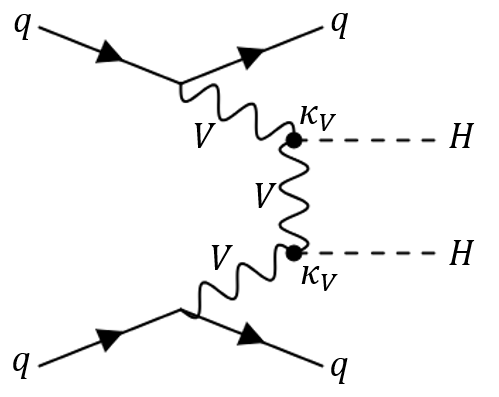}}\hspace{0.2cm}
    \subfloat[e][VBF $\kappa_{2V}$]{\includegraphics[width=0.28\linewidth]{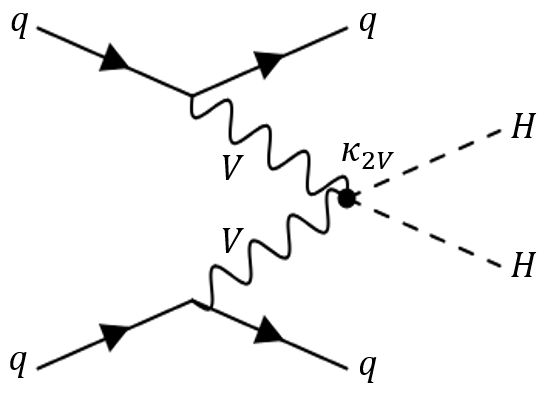}}
    \caption{The leading-order Feynman diagrams for (a–b) gluon–gluon fusion and (c–e) vector boson fusion Higgs boson pair production.}
    \label{fig:feyn}
\end{figure}

At the LHC, the Higgs boson pair production process proceeds mainly through the ggF production mode, mediated by heavy-quark loops, predominantly the top quark. The ggF amplitude receives contributions from two destructively interfering diagrams: the \emph{triangle} diagram, which explicitly depends on the trilinear coupling, and the \emph{box} diagram, which does not. The destructive interference between these two amplitudes makes the total production rate highly sensitive to deviations in the coupling modifier $\kappa_{\lambda}$. A smaller but complementary contribution arises from VBF production, which also depends on $\kappa_{\lambda}$ and, in more general parameterizations, on the Higgs–vector boson couplings $\kappa_V$ and $\kappa_{2V}$. Figure~\ref{fig:feyn} shows the leading-order Feynman diagrams contributing to Higgs boson pair production in both ggF and VBF channels.

In the SM, the ggF mode dominates with a next-to-next-to-leading order (NNLO) cross-section of 
$\sigma_{\text{ggF}}$ = 0.3413 pb at $\sqrt{s}$ = 13.6 TeV and $m_H$ = 125 GeV \cite{CERNYellowReportPageBR}, while the VBF contribution is smaller by nearly two orders of magnitude, 
$\sigma_{\text{VBF}}$ = 0.001874 pb \cite{Dreyer_2020}. Theoretical uncertainties arise primarily from parton distribution functions (PDFs), $\alpha_s$, and QCD scale variations, typically at the 2–5\% level for the VBF process and up to 20\% for ggF \cite{CERNYellowReportPageBR, Dreyer_2018, Dreyer_2020}.

Although small in the SM, the double Higgs boson cross-section can be significantly enhanced in BSM scenarios where new scalar degrees of freedom modify the Higgs potential or introduce resonant production modes. Examples include extended scalar sectors such as two-Higgs-doublet models (2HDM), singlet-extended models (2HDM+S), and supersymmetric frameworks~\cite{Branco:2011iw,Ellwanger:2009dp,Robens:2019kga}. In this work, however, we restrict our study to the non-resonant production regime, where variations in the couplings affect only the overall production rate and event kinematics \cite{Grazzini_2018}.

\begin{figure}[htb]
   \centering
    \subfloat[ggF][ggF]{\includegraphics[width=0.5\linewidth]{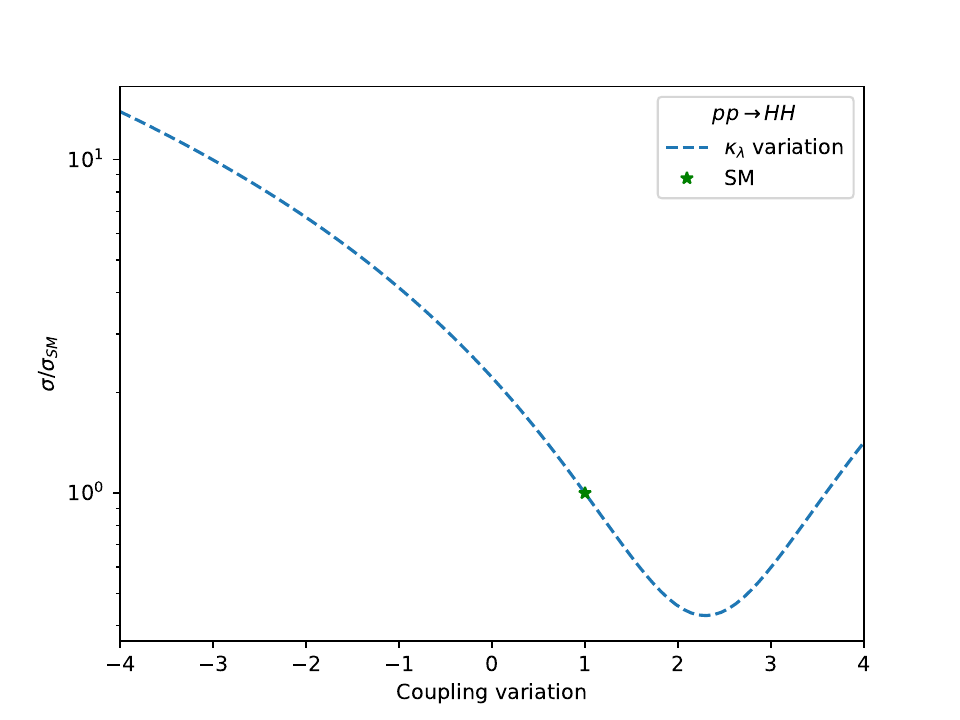}}
    \subfloat[VBF][VBF]{\includegraphics[width=0.5\linewidth]{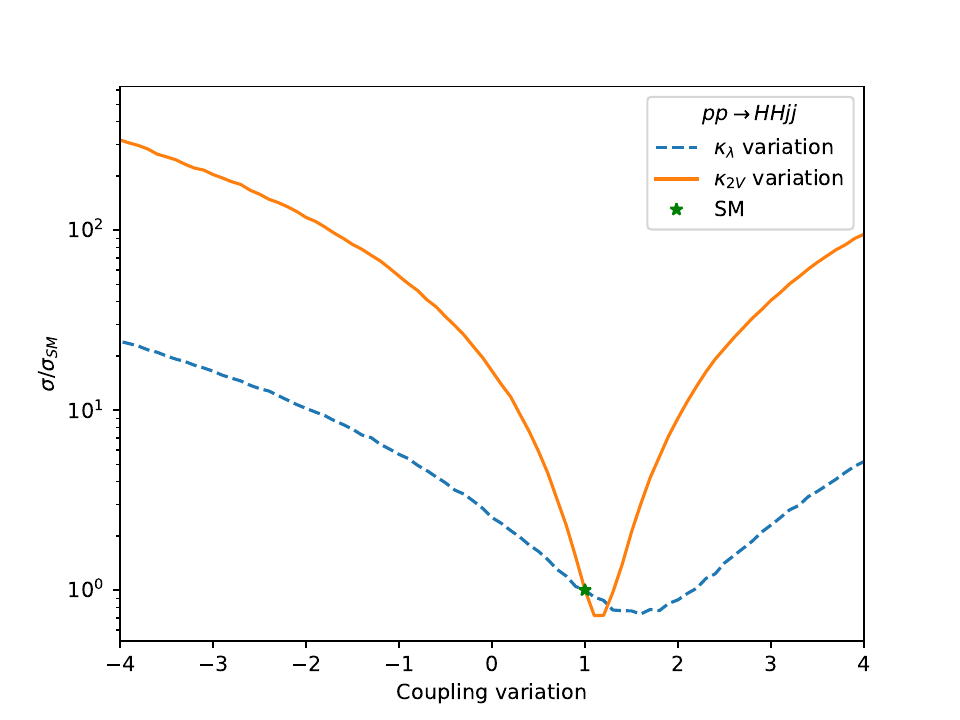}}
   \caption{Variation of (a) ggF and (b) VBF double Higgs boson cross-section as a function of $\kappa_{\lambda}$ and $\kappa_{2V}$. If one parameter varies, the remaining parameters are fixed to 1.}
    \label{fig:xsec_var}
\end{figure}

The dependence of the ggF and VBF cross-sections on $\kappa_{\lambda}$ can be expressed through quadratic parameterizations ~\cite{Dreyer_2020,Belfkir:2025zlx}:

\begin{equation}
    \label{eq:kl}
    \sigma_{\text{ggF}}(\kappa_{\lambda}) = 75.76 - 53.29\,\kappa_{\lambda} + 11.61\,\kappa_{\lambda}^2 \ \text{[fb]}, 
\end{equation}
\begin{equation}
    \label{eq:k2v}
    \sigma_{\text{VBF}}(\kappa_{\lambda}) = 0.0032 - 0.0029\,\kappa_{\lambda} + 0.00093\,\kappa_{\lambda}^2 \ \text{[fb]}.
\end{equation}
The VBF cross-section is generally parameterized including also $\kappa_{2V}$ and $\kappa_{V}$ \cite{Dreyer_2020}. As can be seen in Figure \ref{fig:xsec_var}, when $\kappa_\lambda$ deviates from the SM value, the cross-sections change accordingly demonstrating the sensitivity of these production modes to the Higgs self-coupling and making them important probes for BSM physics.

\section{Monte Carlo simulation}
\label{sec:MC}

This study uses simulated proton–proton collision events at a center-of-mass energy of $\sqrt{s}=13.6~\text{TeV}$ to explore the potential of HyQML in improving the sensitivity to non-resonant Higgs boson pair production in the $b\bar{b}\gamma\gamma$ decay channel. The analysis uses an integrated luminosity of 308 fb$^{-1}$, corresponding to the partial Run-3 + Run-2 dataset collected by the ATLAS detector between 2018 and 2024. The targeted final state consists of two well-identified photons and two $b$-quark tagged jets, originating from $H\to\gamma\gamma$ and $H\to b\bar{b}$ decays. This channel remains one of the most sensitive probes of the Higgs boson self-coupling among all double Higgs boson decay modes \cite{Belfkir:2025zlx}.

The simulated events include both signal and dominant SM backgrounds. The signal processes comprise ggF and VBF Higgs boson pair production, generated with the trilinear Higgs coupling fixed to its SM value. The ggF process is simulated at next-to-leading order (NLO) using \texttt{Powheg-Box~v2}~\cite{Alioli_2010}, including finite top-quark mass effects, while VBF events are generated at leading order (LO) using \texttt{MadGraph5\_aMC@NLO~v3.3.0}~\cite{Alwall_2014}. Both signal samples are normalized to the most recent theoretical cross-section predictions evaluated at NNLO for ggF, and at N3LO QCD plus NLO electroweak accuracy for VBF at Higgs mass $m_H$ = 125 GeV~\cite{Dreyer_2020,Grazzini_2018}.

The background simulation follows the same methodology. Resonant backgrounds arise from single Higgs boson production where $H\to\gamma\gamma$, including also ggF, VBF, associated production with a Z boson (Z($\to b\bar{b}$) H($\to\gamma\gamma$)), and associated production with top quarks ($t\bar{t}$H), are simulated using \texttt{Powheg-Box~v2}. Non-resonant $\gamma\gamma$ + jets background originates from QCD-induced di-photon events accompanied by jets, which can mimic the $b\bar{b}\gamma\gamma$ final state without an intermediate Higgs resonance, is simulated with \texttt{MadGraph5\_aMC@NLO}, with matrix elements including up to two additional partons to model the jet multiplicity spectrum accurately. Table \ref{tab:mc_table} summarizes the simulated processes with respective number of events and cross-sections.
\begin{table}[htb]
    \centering
    \begin{tabular}{lccc}
    \hline\hline
     Process &  Event Generator & Number of events & Cross-section [pb] \\
     \hline
    ggF HH ($\kappa_{\lambda}$ = 1) & Powheg & 50k & 0.3413 \cite{Dreyer_2020}\\
    VBF HH ($\kappa_{\lambda}$ = 1) & MadGraph & 100k & 0.001874 \cite{Dreyer_2020}\\
    \hline 
    ggF Higgs & Powheg & 500k & 52.17 \cite{CERNYellowReportPageBR}\\
    VBF Higgs & Powheg & 550k & 4.075 \cite{CERNYellowReportPageBR}\\
    $qq \to ZH $ & Powheg & 50k & 1.834 $\times 10^{-3}$ \cite{CERNYellowReportPageBR}\\
    $gg \to ZH $ & Powheg & 100k & 3.087 $\times 10^{-4}$ \cite{CERNYellowReportPageBR}\\
    $t\bar{t}$ Higgs & Powheg & 100k & 5.688 $\times 10^{-1}$ \cite{CERNYellowReportPageBR}\\
    \hline 
    $\gamma\gamma $ + jets & MadGraph & 2.5M & 48.1 \\
    \hline\hline
    \end{tabular}
    \caption{Summary of generated signal and background events.}
    \label{tab:mc_table}
\end{table}

All generated samples are processed through \texttt{Pythia~8.186}~\cite{pythia8} for parton showering, hadronization, and underlying-event modeling. The detector response is emulated using the \texttt{Delphes}~\cite{delphes} fast simulation framework configured with an ATLAS detector configuration card. The detector card has been tuned to reproduce the latest ATLAS performance calibrations.

\section{Object definitions and event selections}
\label{sec:selection}

\subsection{Object definitions}

The event topology for the $HH \to b\bar{b}\gamma\gamma$ signal consists of two isolated photons and two $b$-identified jets in the final state. To ensure consistent and accurate object reconstruction, standard Run-3 ATLAS detector definitions are adopted and emulated in the \texttt{Delphes} fast simulation framework.

Photons are reconstructed from energy deposits in the electromagnetic calorimeter using a tower-based clustering algorithm. Each photon candidate is required to have a transverse momentum of $p_T>20~\text{GeV}$ and lie within the pseudorapidity range $|\eta|<2.37$, excluding the calorimeter barrel–endcap transition region $1.37<|\eta|<1.52$, where the detector response deteriorates due to reduced granularity and inactive material~\cite{ATLAS:2022hmt}. To avoid double counting of closely spaced clusters and ensure high isolation efficiency, a self-overlap removal criterion is applied: if two reconstructed photon candidates are separated by less than $\Delta R<0.01$, only the leading (highest-$p_T$) photon is retained. Photon candidates must also satisfy the \emph{Tight} identification working point (WP) emulated with the Run-3 photon identification efficiency at 13.6 TeV~\cite{photonID}.

Jets are reconstructed from calorimeter energy deposits using the anti-$k_t$ algorithm~\cite{Cacciari_2008} with a radius parameter $R=0.4$, as implemented in the \texttt{FastJet} package~\cite{Cacciari:2011ma}. The inputs to the clustering algorithm are calorimeter towers representing the energy flow in the detector. Selected jets must satisfy $p_T>25~\text{GeV}$ and a rapidity $|y|<4.5$, ensuring full containment within the ATLAS calorimeter acceptance. Jets originating from $b$-quarks are identified using a parameterization of the ATLAS $b$-tagging WP at the 85\% efficiency~\cite{btagGN2}. This operating point corresponds to mis-tag rates of approximately 0.17 for $c$-jets and 0.01 for light-flavor jets, allowing realistic modeling of background contamination from mis-identified jets. Only $b$-tagged jets within the inner-detector tracking region ($|\eta|<2.5$) are considered in the analysis.

Leptons (electrons and muons) are reconstructed from the particle-flow track collection. Muon candidates are required to have $p_T>10~\text{GeV}$ and $|\eta|<2.7$, while electrons must satisfy $p_T>10~\text{GeV}$ and $|\eta|<2.47$, excluding candidates in the same calorimeter transition region as photons. These criteria ensure that leptons are well-measured and isolated from hadronic activity, enabling the rejection of events with leptonic signatures from top-quark or electroweak background processes.

Standard object-cleaning criteria are imposed to remove overlaps between reconstructed objects and to ensure isolation of photon and jet candidates. All reconstructed objects are required to be mutually distinct and to pass the default overlap-removal procedure implemented in the \texttt{Delphes} configuration. The resulting set of calibrated and isolated photons, jets, $b$-jets, and leptons forms the input basis for the subsequent event selection and quantum machine-learning classification described later.

\subsection{Event selections}

Events are required to pass a di-photon trigger optimized for the selection of final states containing two energetic photons. The trigger demands that the leading and sub-leading photons have transverse energies exceeding 35~GeV and 25~GeV, respectively. The corresponding trigger efficiency is applied to simulated events using the publicly available measurements derived from ATLAS Run-3 data at $\sqrt{s}=13.6~\text{TeV}$~\cite{trig}. 

Following the selection strategy adopted in previous ATLAS analyses of the $HH\to b\bar{b}\gamma\gamma$ channel~\cite{bbyy_legacy,atlas_results,bbyy_run2}, the two leading photons in each event are required to have an invariant mass within the range $105 < m_{\gamma\gamma} < 160~\text{GeV}$. To maintain a uniform efficiency across the diphoton mass spectrum, dynamic transverse momentum thresholds are imposed, requiring $p_T^{\gamma_1} > 0.35\, m_{\gamma\gamma}$ for the leading photon and $p_T^{\gamma_2} > 0.25\, m_{\gamma\gamma}$ for the sub-leading photon. These conditions prevent biases in the background distribution near the Higgs boson mass peak~\cite{ATLAS:2014euz}. The selected photon pair is used to reconstruct the $H\to\gamma\gamma$ candidate.

The reconstruction of the $H\to b\bar{b}$ candidate is performed using the two leading $b$-tagged jets in the event, both satisfying the selection criteria described in Section~\ref{sec:selection}. Events are required to contain at least two $b$-tagged jets. To suppress background contributions from top-quark processes, particularly $t\bar{t}H$ production, events containing identified electrons or muons are vetoed. In addition, to reduce hadronic contamination from $t\bar{t}H$ and multi-jet processes, the total number of reconstructed central jets in each event is restricted to six or fewer.

Beyond the baseline selection, the topology of the event is further characterized by identifying potential VBF jets. When present, the two highest-$p_T$ jets not associated with the $H\to b\bar{b}$ reconstruction are labeled as VBF-tagged jets. These jets typically correspond to forward-scattered quarks in the VBF process and provide valuable information on VBF event topology. However, the presence of VBF-tagged jets is not mandatory, and events lacking such jets are retained in the inclusive selection. A dedicated VBF event category, which could enhance sensitivity to anomalous couplings such as $\kappa_{2V}$, is not defined in the present study but will be explored in future analyses.

\section{Hybrid-Quantum Machine Learning Model}
\label{sec:hyQML}

This section describes the HyQML model used to enhance the event categorization of the analysis. The proposed HyQML approach combines a classical neural network encoder with a parameterized quantum circuit (PQC), allowing the system to learn both non-linear and quantum-correlated feature representations. The objective is to improve the separation between signal and background events beyond what is achievable with classical XGBoost model or purely classical machine learning selections.

\subsection{Data pre-processing}

Before training the hybrid model, a dedicated data pre-processing pipeline is implemented to standardize the input features and minimize detector-induced asymmetries. In particular, events passing the selection described in Section~\ref{sec:selection} are geometrically rotated in the transverse plane such that the leading photon is aligned with the beam axis. This transformation exploits the cylindrical symmetry of the ATLAS detector in the azimuthal angle $\phi$, effectively removing arbitrary rotational variations and allowing the model to focus on the intrinsic event topology~\cite{Chatham_Strong_2020}. All other reconstructed objects, including the sub-leading photon and the two $b$-jets, are rotated accordingly to preserve their relative spatial configuration. Empirically, this geometric normalization improves the stability of the training process and enhances the model's ability to identify kinematic correlations relevant to Higgs boson pair production.

All the input variables listed in Table \ref{tab:var} are subsequently standardized so that their distributions are centered and scaled to unit variance. This normalization step ensures numerical stability and accelerates convergence during the optimization of both classical and quantum parameters by bringing all observables to comparable magnitudes~\cite{standarz}. The standardized dataset is then randomly partitioned into two statistically independent subsets: 75\% of the events are used for model training, while the remaining 25\% are reserved exclusively for performance evaluation and statistical treatment. The model performance is therefore assessed only on unseen events to ensure an unbiased estimation of generalization capability.

\begin{table}[htb]
    \centering
    \begin{tabular}{ll}
    \hline\hline
        Objects & variables  \\
    \hline
       Photons  & $p_{T}$, $\eta$, $\phi$ \\
       $b$-jets & $p_{T}$, $\eta$, $\phi$ \\
       $\gamma\gamma$ system & $p_{T}$, $\eta$, $\phi$, $m$ \\
       $b\bar{b}$ system& $p_{T}$, $\eta$, $\phi$, $m$  \\
       $b\bar{b}\gamma\gamma$ system &  $p_{T}$, $\eta$, $\phi$, $m$ \\
       VBF-jets & $p_{T}$, $\eta$, $\phi$, $m_{jj}$, $\Delta \eta_{jj}$ \\
     \hline\hline  
    \end{tabular}
    \caption{Input variables for the HyQML model, where $m_{jj}$ is the invariant mass of the two VBF jets and $\Delta \eta_{jj}$ is their $\eta$ separation.}
    \label{tab:var}
\end{table}

Each event is assigned a weight proportional to its generator-level cross-section, ensuring that the composition of the training sample reflects the expected yield of signal and background processes in data. To further mitigate the strong class imbalance between the simulated signal and backgrounds, an additional weighting factor is applied based on class frequencies, computed using the \texttt{compute\_sample\_weight} function from the \texttt{scikit-learn} package~\cite{scikitlearn}. The total event weight is therefore defined as the product of the generator-level cross-section weight and the class weight. This strategy ensures that the hybrid model remains sensitive to both signal and background contributions during training, avoiding bias toward the more abundant background events. 

\subsection{HyQML model classifier}

The proposed classification model employs a HyQML model designed to integrate classical feature encoding with quantum parameterized circuits in a unified end-to-end training scheme \cite{AitHaddou:2025lei,haug2022natural}. The model combines two key components: 

\begin{enumerate}
    \item \textit{Meta-Parameter Mapping Network} that generates adaptive quantum circuit parameters conditioned on the event kinematics.
    \item \textit{Parameterized Quantum Circuit} that performs quantum feature transformation and outputs expectation values used for event classification.
\end{enumerate}

This hybrid architecture is implemented using \texttt{PennyLane} interfaced with \texttt{PyTorch} for automatic differentiation \cite{Bergholm:2018cyq,pytorch}.

\subsubsection{\textbf{Meta-Parameter Mapping Network}}

The meta-parameter mapping network (\texttt{MetaParamMapNet}) is a fully connected feed-forward neural network that maps the classical kinematic features of each event to the trainable parameters of the PQC. By conditioning the PQC on event-by-event information, this network enables the quantum model to adapt its state preparation and entanglement structure to the underlying physics, thereby enhancing its expressive power for signal to background discrimination.

For an input vector of dimension $d_{\text{in}}$, the network outputs $d_{\theta}$ rotation angles that parameterize the single-qubit gates in the PQC:
\[
d_{\theta} = N_{\text{qubits}} \times N_{\text{layers}} \times 3,
\]
where the factor of 3 corresponds to the independent $\mathrm{R}_X$, $\mathrm{R}_Y$, and $\mathrm{R}_Z$ rotations applied to each qubit in every layer.

The network consists of three hidden layers of 64 neurons, each followed by ReLU activation and a normalization layer \cite{relu}. This architecture provides sufficient non-linearity to model complex correlations in the input features while maintaining stable gradients during training. The output of the encoder can be expressed as:
\[
h = \text{LayerNorm}\!\left(\text{ReLU}\!\left(W_2\, \text{ReLU}(W_1 x + b_1) + b_2\right)\right),
\]
which is subsequently projected onto the $d_{\theta}$-dimensional parameter space of the PQC.

Each PQC parameter $\theta_i$ in the proposed HyQML model is produced by a dedicated linear head within the MetaParamMapNet. After the shared encoder processes the input variables through several nonlinear layers and produces a latent representation $h$, this vector is passed to a group of small, independent linear sub-networks, or "heads" each head corresponds to one learnable parameter of the PQC. Specifically, the i-th head applies a simple linear transformation with its own weight vector and bias term to generate a single scalar value $\Delta\theta_i$ :
\[
\Delta\theta_i = W_i^{(\text{head})} h + b_i^{(\text{head})},
\]
and combined with a learnable base parameter vector $\theta_0$ as
\[
\theta = \theta_0 + \Delta\theta,
\]
which is then squashed into the range $[-\pi, \pi]$ using a sigmoid-based rescaling.  
This parameterization allows the network to produce a distinct set of quantum rotation angles for every event, effectively learning a data-dependent embedding in the quantum circuit parameter space.

\subsubsection{\textbf{Parameterized Quantum Circuit }}
The parameterized quantum circuit in the proposed HyQML framework is implemented as a hardware-efficient ansatz with $N_{\text{qubits}} = 4$ and $N_{\text{layers}} = 4$. A hardware-efficient ansatz means the circuit is designed to be easily realizable on near-term quantum hardware, which uses standard rotation and entangling gates that can be implemented on most quantum processors. This structure strikes a balance between expressiveness and practicality, providing sufficient parameterized gates to capture complex data relationships while maintaining a shallow enough circuit to remain trainable and computationally efficient.

Before quantum computation begins, the input features from the classical dataset must be embedded into a quantum state. In this model, we use amplitude embedding, which encodes the classical feature vector $x$ directly into the amplitude coefficients of a quantum state:
\[
|\psi(x)\rangle = \text{AmplitudeEmbedding}(x),
\]
which normalizes and loads the input vector into the quantum amplitude space of dimension $2^{N_{\text{qubits}}}$.  
Each layer of the quantum circuit performs a series of qubit rotations followed by entangling CNOT operations \cite{Bataille:2020zzg}. The rotations $R_X$, $R_Y$, and $R_Z$ are parameterized by the trainable angles $\theta_{l,i,1}$, $\theta_{l,i,2}$, and $\theta_{l,i,3}$ for qubit $i$ in layer $l$. These rotations allow each qubit to explore the full Bloch sphere, providing a flexible basis for representing complex transformations of the encoded input state. After all qubits in a layer are rotated, entanglement is introduced through a sequence of CNOT gates arranged in a ring topology, where each qubit acts as a control for the next one. This structure ensures that correlations propagate throughout the system, enabling both local and global quantum interactions. This operation is expressed as: 

% COMMENT ADDED BY MARWAN :

\[
U(\theta) = 
\prod_{i=1}^{N_{\text{qubits}}} \text{CNOT}(i, (i+1)\bmod N_{\text{qubits}}).
\prod_{l=1}^{N_{\text{layers}}}
\left[
\prod_{i=1}^{N_{\text{qubits}}}
R_Z^{(i)}(\theta_{l,i,3}) R_Y^{(i)}(\theta_{l,i,2}) R_X^{(i)}(\theta_{l,i,1})
\right]
\]

After that, the expectation values of the Pauli-Z operator are measured on each qubit from the quantum feature vector that represents the output of the parameterized quantum circuit. For the $j$-th qubit, the expectation value is defined as:
\[
z_j = \langle \psi(x, \theta) | Z_j | \psi(x, \theta) \rangle, \quad j = 1,\dots,N_{\text{qubit}}.
\]

These measured values capture the quantum correlations and interference effects generated within the circuit, encoding the processed information from the embedded input event kinematics. The resulting vector $\mathbf{z} = [z_1, z_2, \ldots, z_{N_{\text{qubits}}}]^{\top}$ is then passed to a classical linear output layer, which maps the \(N_{\text{qubits}}\) expectation values to \(N_{\text{classes}} = 2\) logits corresponding to the classification outputs. This final step converts the quantum features into classical decision variables, completing the hybrid quantum--classical learning pipeline.

\begin{figure}[htb]
    \centering
    \includegraphics[width=1.0\linewidth]{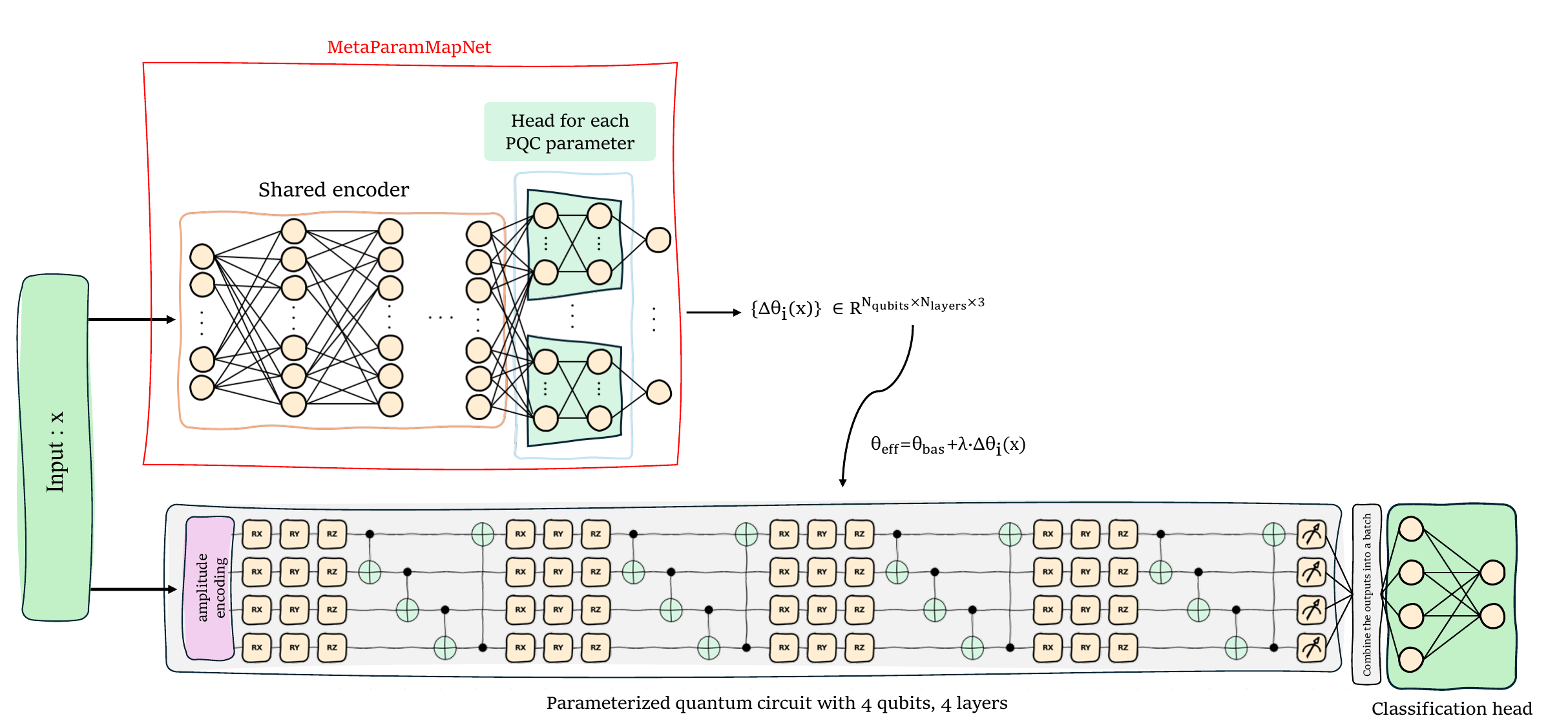}
    \caption{A schematic view of the hybrid quantum machine learning model.}
    \label{fig:model}
\end{figure}

A schematic view of the full hybrid quantum machine learning model is shown in Figure \ref{fig:model}.

\subsection{Hybrid Training Procedure}

The hybrid training objective defines how the classical and quantum components of the model are combined and optimized together. In this framework, the overall function $f(x)$ represents the hybrid model that takes a classical input and processes it through both the meta-network and the PQC. The quantum circuit then computes the expectation values of observables, which form the quantum feature vector. These values are passed to a classical linear layer that outputs the final prediction. The model demonstrates that the quantum circuit generates learned features that are linearly combined to produce the classification output. The hybrid model can be simplified as:
\[
f(x) = \text{Linear}\!\left( \mathbb{E}_{\text{PQC}}(\theta(x)) \right).
\]
where $\theta(x)$ are the parameters predicted by the meta-model.  
A scalar coefficient $\lambda$ regulates the contribution of the meta-network to the final circuit parameters \cite{AitHaddou:2025lei}, allowing gradual control between static and data-dependent parameterization:
\[
\theta_{\text{eff}} = \theta_{\text{base}} + \lambda \, \Delta\theta(x).
\]

For different values of the hyperparameter $\lambda$, the full model is trained using the standard cross-entropy loss function, which measures the difference between the predicted and true class probabilities, which is defined as:
\[
\mathcal{L} = - \sum_i w_i \, y_i \log \hat{y}_i,
\]
where $w_i$ is per-event weight. Training is performed using the Adam optimizer with a learning rate of $10^{-3}$ for 20 epochs, and gradient clipping on the PQC parameters $\theta$ to prevent instability from large updates.

A meta-learning stage precedes the hybrid model training, during which the meta-parameter network is optimized independently to minimize the condition number of the Fubini--Study metric $g_{ij}$ of the PQC~\cite{hospedales2021meta}. This metric characterizes the local geometry of the quantum parameter space and provides a diagnostic of barren-plateau behavior \cite {Larocca:2024plh}. The meta-objective is defined as:
\[
\mathcal{L}_{\text{meta}} = \left\langle \log \kappa(g_{ij}) \right\rangle,
\]
where $\kappa(g_{ij})$ denotes the condition number of the metric tensor. Minimizing $\mathcal{L}_{\text{meta}}$ improves the expressivity and trainability of the PQC by promoting well-conditioned gradients and mitigating vanishing-gradient regimes~\cite{AitHaddou:2025lei}.

The meta-network is trained using a small batch of randomly sampled events and optimized with the Adam optimizer until convergence. Once this meta-learning phase is completed, the meta-network weights are frozen and remain fixed throughout the subsequent quantum-circuit training stage, ensuring that the PQC is trained on a stable and geometry-optimized parameterization.

\subsection{Hybrid Model Performance}

\begin{figure}[htb]
    \centering
    \includegraphics[width=0.7\linewidth]{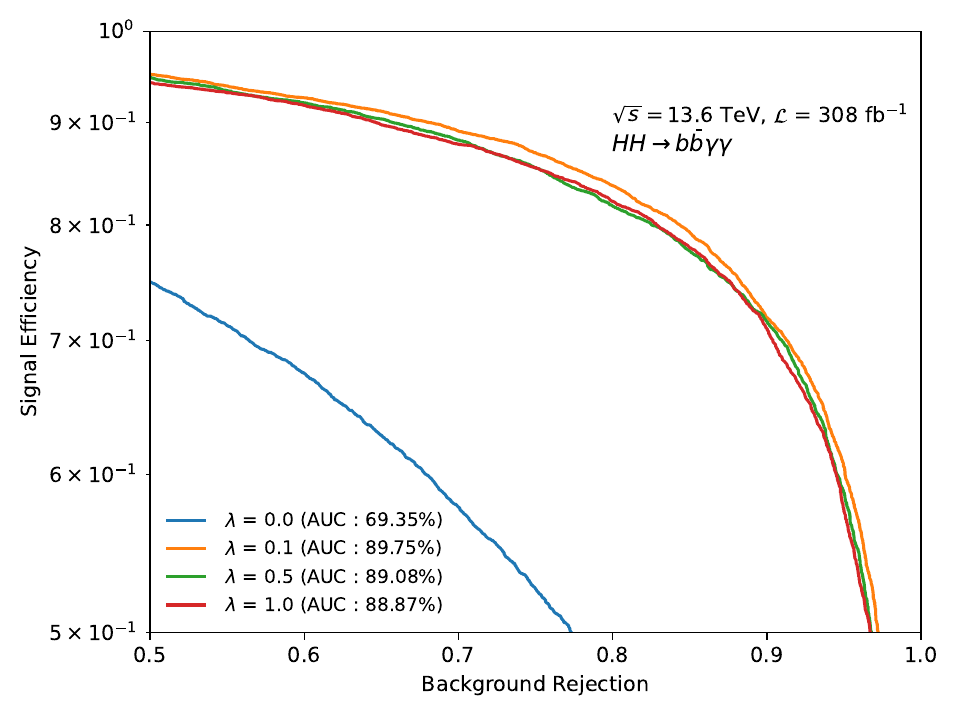}
    \caption{Weighted ROC curve for the HyQML classifier for different $\lambda$ values. The ROC curve is computed using event weights on the test dataset.}
    \label{fig:ROC}
\end{figure}

The performance of the trained quantum classifier is evaluated using the Receiver Operating Characteristic (ROC) curve, which quantifies the trade-off between signal efficiency (true positive rate) and background rejection (false positive rate) as the decision threshold on the model output score is varied. The area under the ROC curve (AUC) serves as a global measure of classification performance. Figure~\ref{fig:ROC} shows the ROC curve obtained for the hybrid quantum model for different hyperparameter $\lambda$ values. It is found that the $\lambda$ = 0.1 provides a better discrimination power between the $HH \to b\bar{b}\gamma\gamma$ signal and the dominant background processes. The case $\lambda = 0$ corresponds to a pure quantum circuit (Pure-QML), which shows very low performance compared to the HyQML model. The rest of the results are shown for $\lambda = 0.1$.

\begin{figure}[htb]
   \centering
    \subfloat[][First epoch]{\includegraphics[width=0.5\linewidth]{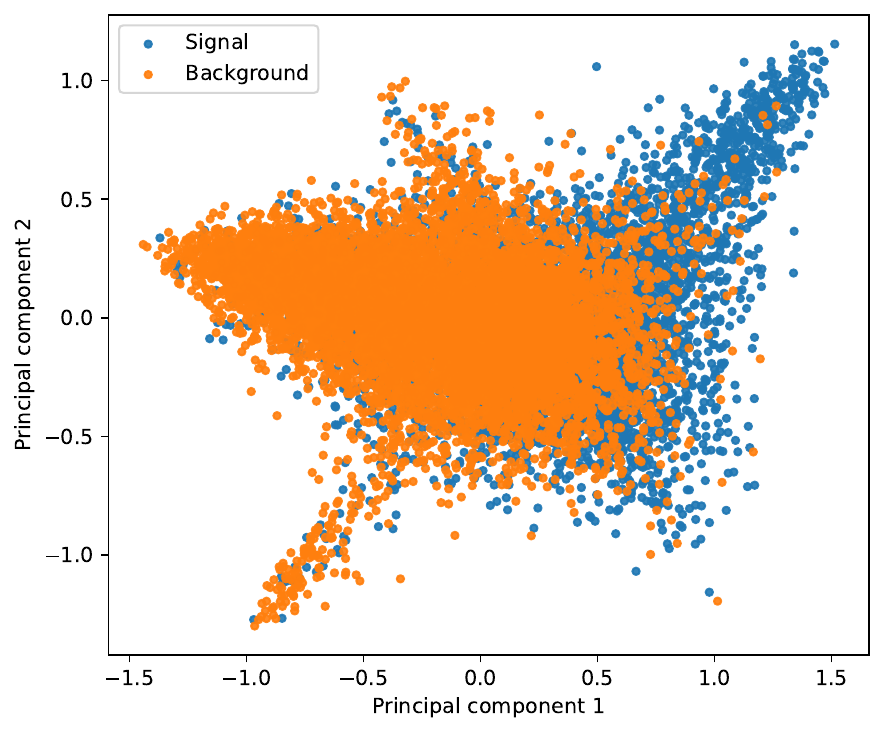}}
    \subfloat[][Last epoch]{\includegraphics[width=0.5\linewidth]{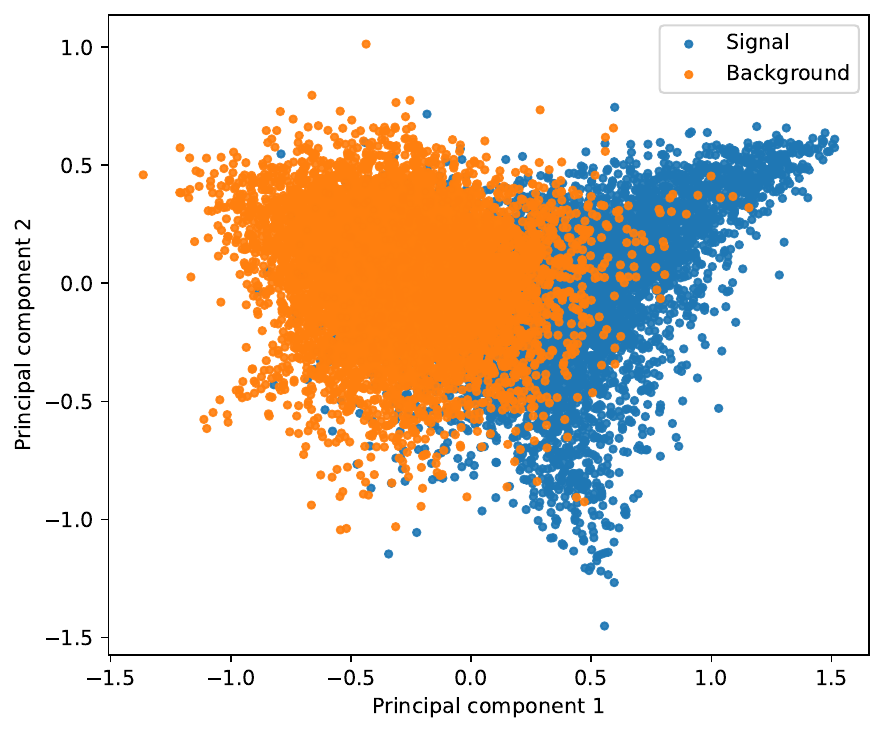}}
   \caption{Two-dimensional projection of the PQC expectation values for the first and the last epochs for $\lambda$ = 0.1. }
    \label{fig:pca}
\end{figure}

To gain insight into the internal representations learned by the hybrid quantum model, we monitor the evolution of the PQC output during training. At each epoch, the expectation values of the PQC measurements are collected and projected onto a two-dimensional space using Principal Component Analysis (PCA) \cite{PCA}. Figure \ref{fig:pca} shows the PCA projection between the first and the last epoch in the training of the hybrid quantum model.
These projections provide an interpretable visualization of the quantum latent space, allowing one to monitor the progressive separation between signal and background distributions. At an early stage of training, the signal and background samples occupy largely overlapping regions in the latent space, indicating that the PQC has not yet learned a meaningful transformation of the input kinematic features. At a later training epoch, a clear organization of the latent space emerges: the signal and background distributions begin to separate into distinguishable clusters, and the points align along a more coherent low-dimensional manifold. The latent topology also undergoes a noticeable rotation relative to the early-epoch projection, reflecting the progressive restructuring of the quantum feature space as the model learns. The reduced overlap between the two classes demonstrates that the PQC increasingly captures discriminative patterns relevant for classification. This diagnostic step reveals that the HyQML model learns structured manifolds in the quantum feature space, with a gradual increase in class separation and rotation in the latent topology over successive epochs. 

\section{Results}
\label{sec:results}
\begin{figure}[htb] 
    \centering
    \includegraphics[width=0.7\linewidth]{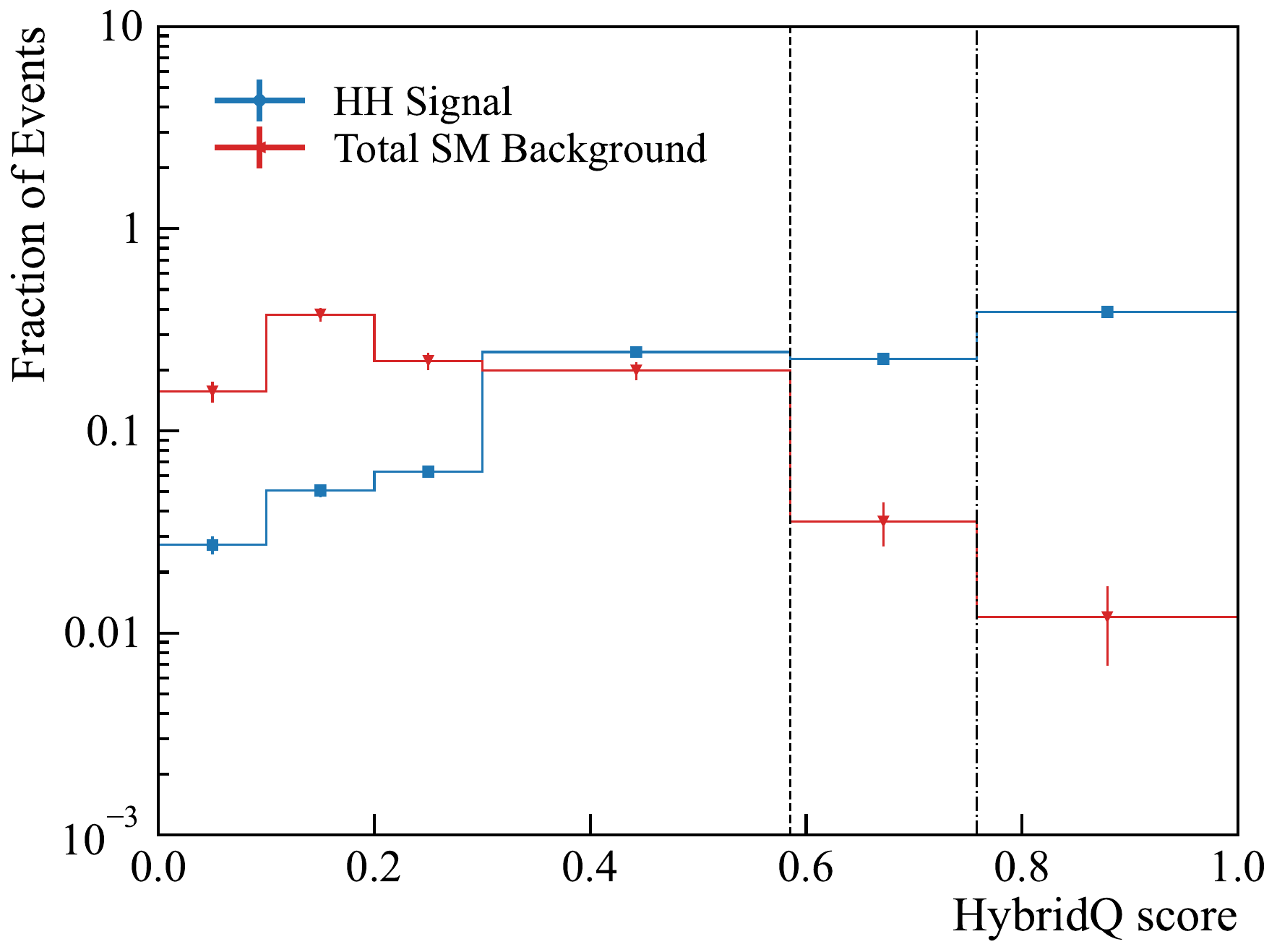}
    \caption{Distribution of the HyQML score. The vertical dashed line indicates the optimized bins threshold.}
    \label{fig:score}
\end{figure}

The goal of this analysis is to enhance the sensitivity to non-resonant Higgs boson pair production using a HyQML classifier designed to discriminate between signal and background events. The model combines quantum feature mapping with classical optimization to exploit both non-linear correlations and quantum-induced entanglement patterns that are otherwise inaccessible to classical architectures. To quantify the statistical sensitivity gain, the output distribution of the HyQML classifier is divided into optimized score regions corresponding to increasing signal purity. In particular, the signal-enriched region is defined for classifier scores above 0.5, where the quantum model exhibits the strongest separation between signal and background. The optimized regions are required to maximize the expected statistical significance computed using the Asimov approximation to the profile likelihood ratio~\cite{Cowan:2010js} defined as: 

\begin{equation*}
    Z = \sqrt{2 \times ((s + b) \times \ln(1 + s / b) - s)},
\end{equation*}
where $s$ is the total signal yield and $b$ corresponds to the total background yield. In each defined bin a requirement of at least one background event is imposed. The lower-score region, containing predominantly background events, is grouped into a single inclusive bin as it contributes marginally to the overall sensitivity. Figure~\ref{fig:score} presents the HyQML output score distribution for both signal and total SM background in the signal region defined before. The vertical dashed line marks the optimized threshold separating the signal-enriched and inclusive regions. A clear distinction between the two classes is visible, highlighting the HyQML model's ability to capture non-trivial event correlations in the $b\bar{b}\gamma\gamma$ final state.

Due to the limited size of the simulated continuum $\gamma\gamma$+jets background, the effective event statistics are low in high-score regions. Consequently, bins with few high-score background events may have large statistical fluctuations. To assess the robustness of the result, the statistical interpretation is therefore performed with a background uncertainty of 10\% and 50\%. In both cases, the HyQML model maintains stable performance and consistent behavior, confirming that the hybrid quantum approach generalizes well even in low-statistics regimes. The performance of the HyQML is also compared to a pure-QML ($\lambda$  = 0) and the XGBoost model presented in the Ref~\cite{Belfkir:2025zlx} with 10\% background uncertainty.

The sensitivity of the analysis is quantified through two key statistical metrics:  
the expected discovery significance and the 95\% confidence level (CL) upper exclusion limit on the Higgs boson pair production cross-section~\cite{CL}.  
Both quantities are obtained from a binned likelihood fit to the HyQML classifier output distribution using the optimized binning.

The statistical inference is performed with the \texttt{pyhf} package~\cite{Heinrich:2021gyp}, which implements the profile-likelihood formalism.  
For a given signal strength parameter $\mu \ge 0$ defined as,
\[
\mu = \frac{\sigma(pp \to HH)}{\sigma_{\text{SM}}(pp \to HH)},
\]
such that $\mu=1$ corresponds to the SM expectation, and a set of nuisance parameters $\boldsymbol{\theta}$ describing uncertainties corresponding in this analysis to 10\% and 50\%, the likelihood function is defined as
\[
\mathcal{L}(\mu,\boldsymbol{\theta}) =
\prod_{i=1}^{N_{\text{bins}}}
\operatorname{Pois}\!\left(n_i \,\middle|\, \mu\,s_i(\boldsymbol{\theta}) + b_i(\boldsymbol{\theta})\right)
\times
\prod_{k=1}^{N_{\text{nuis}}} \pi_k(\theta_k),
\]
where $n_i$ is the observed yield in bin $i$, while $s_i$ and $b_i$ are the expected signal and background yields, respectively. $n_i$ is defined as the sum of the background yield $b_i$ and the SM ($\kappa_{\lambda} = 1$) expected signal yield $s_i$. 
The constraint terms $\pi_k(\theta_k)$ encode only the background normalization uncertainty of 10\% and 50\% and is implemented as log-normal distributions.

The profile-likelihood ratio is defined as
\[
\lambda(\mu) = 
\frac{\mathcal{L}(\mu,\hat{\hat{\boldsymbol{\theta}}}_{\mu})}
     {\mathcal{L}(\hat{\mu},\hat{\boldsymbol{\theta}})},
\]
where $(\hat{\mu},\hat{\boldsymbol{\theta}})$ are the unconditional maximum-likelihood estimators (MLEs), and $\hat{\hat{\boldsymbol{\theta}}}_{\mu}$ are the conditional MLEs for a fixed signal strength $\mu$.  
The corresponding test statistic for upper limits is
\[
q_{\mu} =
\begin{cases}
-2\ln \lambda(\mu), & 0 \le \hat{\mu} \le \mu, \\
0, & \text{otherwise.}
\end{cases}
\]
The discovery test statistic is obtained by setting $\mu=0$,
\[
q_{0} =
\begin{cases}
-2\ln \lambda(0), & \hat{\mu} \ge 0, \\
0, & \text{otherwise.}
\end{cases}
\]

Under the asymptotic approximation, the one-sided discovery significance is given by~\cite{Cowan:2010js}
\[
Z = \sqrt{q_{0}},
\]
evaluated on an Asimov dataset generated under the signal-plus-background hypothesis ($\mu=1$).  
Similarly, the 95\% CL upper limit on the signal strength $\mu$ is obtained from the modified frequentist $CL_s$ criterion,
\[
CL_s(\mu) = \frac{p(q_{\mu}|\mu)}{1 - p(q_{0}|0)},
\qquad
\text{solve for } \mu \text{ such that } CL_s(\mu) = 0.05.
\]
The same likelihood fit to the HyQML classifier output is used to extract both the expected discovery significance and the 95\% CL upper limit.  

Table \ref{tab:sig} summarizes the expected discovery significance obtained with the fit for both 10\% and 50\% background uncertainties compared to the pure-QML and classical XGBoost, which is found to be consistent with the values reported by the ATLAS experiment \cite{atlas_results}. A factor of two improvement is achieved with the HyQML over the pure-QML in the expected discovery significance.

\begin{table}[htb]
    \centering
    \begin{tabular}{lcccc}
    \hline \hline
         & HyQML (10\% sys.) & HyQML (50\% sys.) & Pure-QML ($\lambda$ = 0) & XGBoost \\
    \hline     
     Significance   &  1.41 & 1.12 & 0.65 & 1.09\\
    \hline\hline
    \end{tabular}
    \caption{Expected discovery significance for both 10\% and 50\% 
background uncertainties compared to the pure-QML and the XGBoost.}
    \label{tab:sig}
\end{table}

\begin{figure}[htb]
    \centering
    \includegraphics[width=0.7\linewidth]{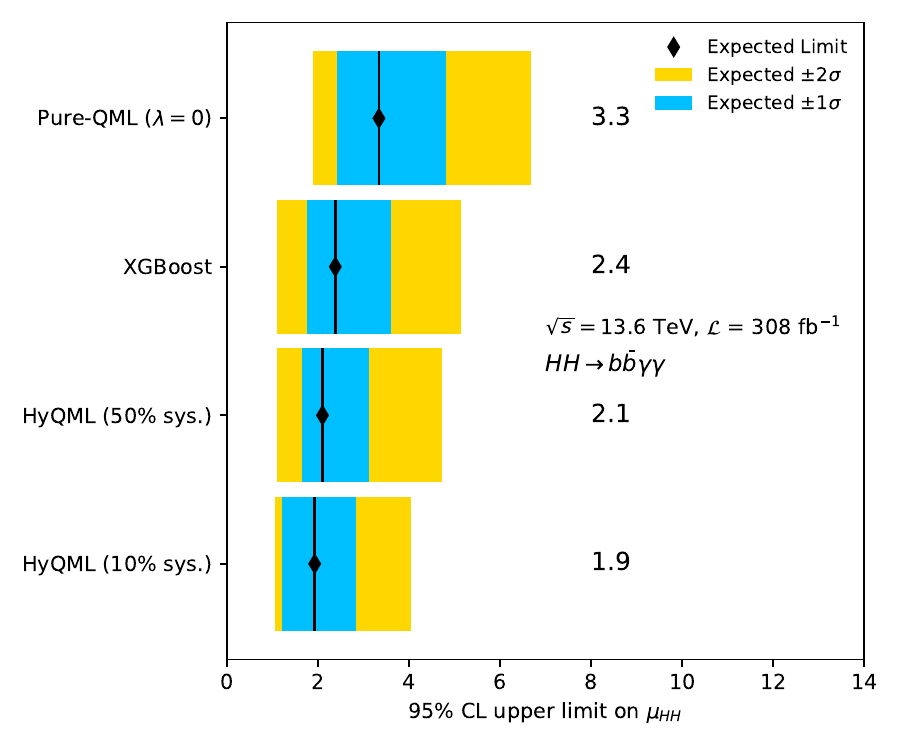}
    \caption{Expected 95\% CL upper limits on the signal strength $\mu_{HH}$ under the 10\%, 50\% background normalization uncertainties conditions and pure quantum model. The shaded bands represent the $\pm1\sigma$ and $\pm2\sigma$ uncertainty intervals.}
    \label{fig:limit}
\end{figure}

Figure~\ref{fig:limit} shows the expected 95\%~CL upper limits on the $\mu_{HH}$ for the two background normalization uncertainties hypotheses and the pure quantum model. Assuming a total background normalization uncertainty of 10\%, the expected limit is approximately \(1.9\times\sigma_{\text{SM}}\). When the uncertainty is increased to 50\%, the limit relaxes slightly to about \(2.1\times\sigma_{\text{SM}}\). Despite this increase, the results remain consistent within uncertainties, demonstrating the robustness of the HyQML approach and its ability to maintain stable performance even under limited-statistics conditions. The HyQML demonstrates a factor of 1.75 improvement in the expected 95\%~CL upper limit compared to the pure quantum model architecture and achieves a 21\% improvement in the upper limit compared to a classical XGBoost.

Despite being based on a simplified detector simulation, the quantum-enhanced classifier achieves improvement compared to classical method used in the latest ATLAS results \cite{atlas_results}. The ATLAS collaboration reports an observed (expected) upper limits on the Higgs boson pair production cross-section of 3.8 (3.7) times the SM prediction. Even though a direct comparison is not optimal, the proposed HyQML achieves a significant improvement.  The performance gain is attributed to the hybrid quantum model's ability to encode correlations in a higher-dimensional feature space through non-classical operations, enabling more efficient separation of signal and background events.

\begin{figure}[htb] 
    \centering
    \subfloat[$\kappa_{\lambda}$][$\kappa_{\lambda}$]{\includegraphics[width=0.5\linewidth]{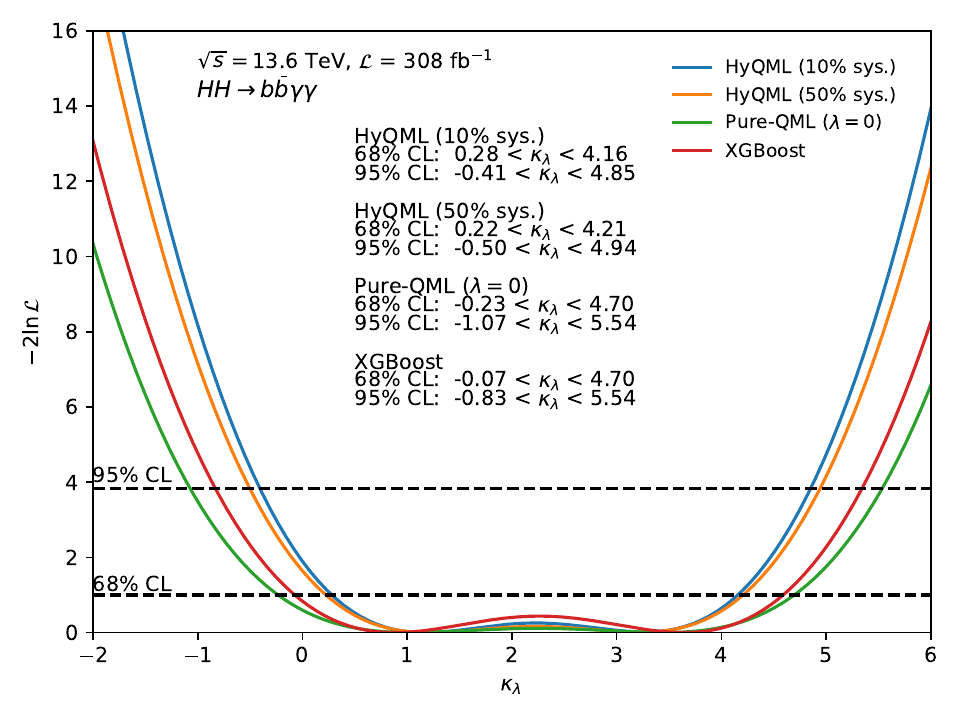}}
    \subfloat[$\kappa_{2V}$][$\kappa_{2V}$]{\includegraphics[width=0.5\linewidth]{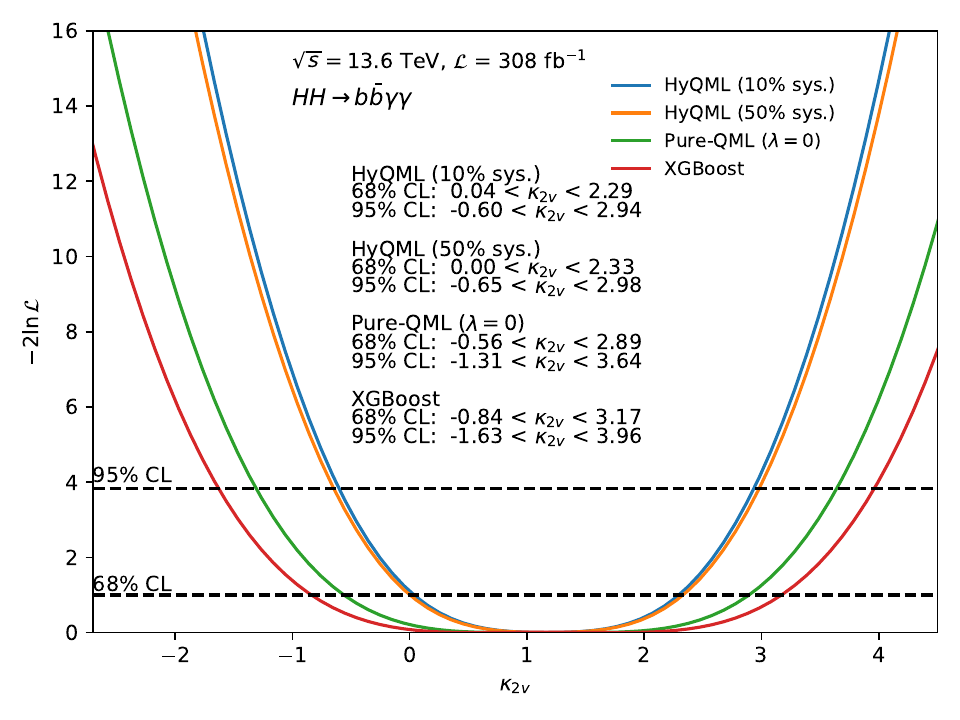}}
    \caption{Negative log-likelihood as a function of the trilinear Higgs self-coupling modifier \(\kappa_{\lambda}\) (a) and \(\kappa_{2V}\) (b) for the 10\%  (blue), 50\% (orange) background uncertainties, the pure-QML (green) and the XGBoost model (red). Dashed lines indicate the 68\% and 95\% confidence level intervals.}
    \label{fig:kl_k2v_scan}
\end{figure}

Although the present study is not explicitly optimized for measuring the trilinear Higgs boson self-coupling modifier, \(\kappa_{\lambda}\), a one-dimensional profile-likelihood scan is performed to estimate the 68\% and 95\% confidence intervals in both scenarios. In this procedure, only \(\kappa_{\lambda}\) is treated as a free parameter while all other couplings are fixed to their SM values. The scan accounts only for the dependence of the total production cross-section on \(\kappa_{\lambda}\) as defined by Equation~\ref{eq:kl}, without including potential kinematic shape variations. Figure~\ref{fig:kl_k2v_scan}(a) shows the negative log-likelihood profile obtained with the 10\%, 50\% background normalization uncertainties and the pure quantum model. From the likelihood scan, the HyQML analysis yields an expected 95\% (68\%) confidence interval of approximately 
\[
\kappa_{\lambda} \in [-0.4,\ 4.9] \ \ ([0.3,\ 4.2]),
\]
which is consistent with the latest experimental constraints of \([-1.7,\ 6.6]\) (95\%) and \([-0.4,\ 5.1]\) (68\%)~\cite{atlas_results} and demonstrates a significant improvement over both the pure quantum architecture and classical XGBoost model. The comparison underscores the potential of quantum-enhanced methods to deliver precision measurements of the Higgs boson self-coupling in future high-luminosity LHC analyses and beyond.

\begin{figure}[htb]
    \centering
    \includegraphics[width=0.7\linewidth]{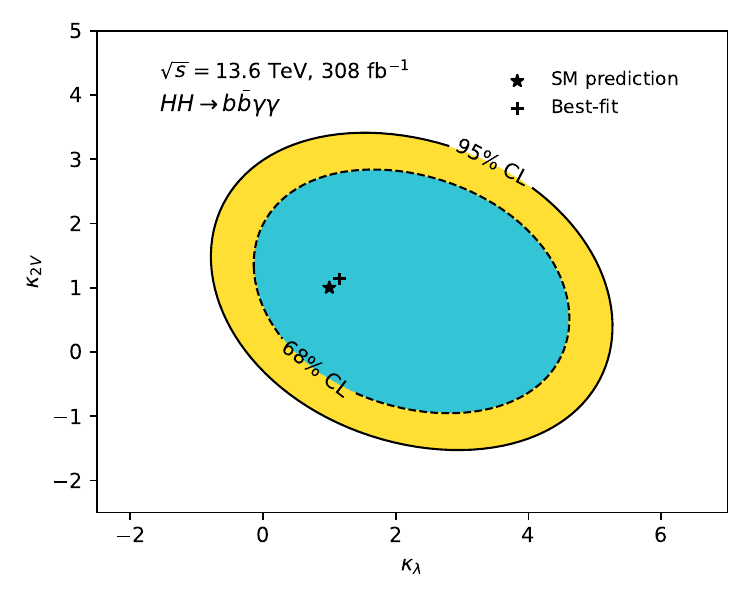}
    \caption{Contours at 68\% CL (dashed line)
and 95\% CL (solid line) in the ($\kappa_{\lambda}$, $\kappa_{2V}$) parameter space. The SM prediction ($\kappa_{\lambda}$  = 1, $\kappa_{2V}$ = 1) is indicated by a star, while the $+$ corresponds to the best-fit values.}
    \label{fig:2D_scan}
\end{figure}

A similar one-dimensional scan is performed for the $\kappa_{2V}$ coupling, which parameterizes deviations in the quartic $VVHH$ interaction entering the VBF Higgs pair production process (Equation~\ref{eq:k2v}). As shown in Figure~\ref{fig:kl_k2v_scan}(b), the HyQML model displays a well-defined likelihood minimum near the SM expectation $\kappa_{2V}=1$, with symmetric confidence intervals at 68\% and 95\% CL. The expected sensitivity obtained from this analysis corresponds to
\[
\kappa_{2V} \in [-0.6,\ 2.9] \ \ ([0.0,\ 2.3]),
\]
We must stress that this analysis does not include a dedicated VBF-category resulting of the lower sensitivity. These results confirm that the hybrid quantum framework maintains robust performance across different coupling hypotheses, providing consistent constraints on both $\kappa_{\lambda}$ and $\kappa_{2V}$ without the need for explicit retraining under new coupling scenarios. In addition, a two-dimensional likelihood fit is performed allowing both $\kappa_{\lambda}$ and $\kappa_{2V}$ to vary. Figure \ref{fig:2D_scan} shows the 95\% and 68\% CL contours with the best-fit values of $\kappa_{\lambda}$ and $\kappa_{2V}$.

Overall, the HyQML training procedure demonstrates a significant improvement over both a pure-QML with $\lambda$ = 0 and a classical XGBoost model. By learning event topology directly from multidimensional kinematic correlations, the quantum-assisted model retains strong discriminating power while generalizing effectively to BSM coupling variations. These results strongly highlight the potential of quantum-enhanced methods as competitive tools for precision Higgs-sector measurements at the LHC and future high-luminosity upgrades.

\section{Conclusion}
\label{sec:conclusion}

In this work, we proposed a HyQML algorithm to enhance the sensitivity of double Higgs boson searches at the LHC. The analysis focuses on the $HH \to b\bar{b}\gamma\gamma$ final state, which, despite its small branching ratio, offers a clean experimental signature and excellent mass resolution. The proposed framework combines parameterized quantum circuits with a classical neural meta-model that dynamically conditions quantum parameters on event-level features, thereby integrating quantum feature representations with the optimization stability of classical learning. This hybrid design bridges the gap between classical and quantum approaches, exploiting high-dimensional correlations that are difficult to capture with traditional models.

The hybrid model outperforms a purely quantum architecture, achieving an improvement of approximately 27\% in the area under the ROC curve (AUC). Furthermore, the HyQML framework yields an expected 95\%~CL upper limit on the non-resonant di-Higgs production cross-section of \(1.9\times\sigma_{\text{SM}}\) assuming a 10\% background normalization uncertainty, and \(2.1\times\sigma_{\text{SM}}\) when the uncertainty is increased to 50\% leading to almost a factor-of-two improvement compared to the pure-QML model and a 21\% improvement with respect to an XGBoost model. These results demonstrate the robustness of the hybrid quantum–classical approach, maintaining stable performance even under large systematic variations and limited training statistics. Likelihood scans of the Higgs self-coupling modifier $\kappa_{\lambda}$ and the quartic vector-boson–Higgs coupling $\kappa_{2V}$ show confidence intervals consistent with recent ATLAS measurements, confirming the model's reliability across different coupling hypotheses.

Overall, this study demonstrates that quantum-assisted learning can achieve strong discriminating power in realistic collider analyses, and with the rapid progress in quantum hardware—particularly improvements in qubit fidelity, circuit depth, and noise mitigation—hybrid quantum models hold significant promise for future searches of new physics at the LHC.

\begin{acknowledgments}
This work is supported by the United Arab Emirates University (UAEU) Start-Up Grant No 12S157. The authors gratefully thank the AI and Robotics Lab of United Arab Emirates University for offering computing facilities including HPC and DGX1 for MC simulation and ML training.
\end{acknowledgments}

\section*{Data and Code Availability}
The datasets and code used in this analysis can be provided by the corresponding author upon reasonable request.\\

\bibliography{apssamp}%
\end{document}